# Soft Manifold Dynamics Behind Negative Thermal Expansion

Z. Schlesinger[1], J. A. Rosen[1], J. N. Hancock[2], and A. P. Ramirez[3]
[1]Physics Department, University of California Santa Cruz, Santa Cruz, CA 95064, USA
[2]Geballe Laboratory for Advanced Materials and SSRL, Stanford University, Palo Alto, CA 95000
[3]Bell Laboratories, Alcatel-Lucent, 600 Mountain Ave, Murray Hill, NJ 07974

Minimal models are developed to examine the origin of large negative thermal expansion (NTE) in under-constrained systems. The dynamics of these models reveals how underconstraint can organize a thermodynamically extensive manifold of low-energy modes which not only drives NTE but extends across the Brilloun zone. Mixing of twist and translation in the eigenvectors of these modes, for which in $ZrW_2O_8$ there is evidence from infrared and neutron scattering measurements, emerges naturally in our model as a signature of the dynamics of underconstraint.

Materials generally expand when heated. The best-known exceptions are found in systems undergoing structural phase transitions, such as the freezing of $H_2O$, where *negative* thermal expansion (NTE) can be large but is restricted to a narrow temperature range. Other examples, such as silicon, possess low NTE at cryogenic temperatures driven by a negative Gruneisen constant for the acoustic modes – the so-called "guitar-string" model[1]. There are, however, a few maverick materials that exhibit large NTE over a wide temperature range, the most exceptional being $ZrW_2O_8$ which contracts isotropically with $\alpha = (1/L)\partial L/\partial T \sim -10^{-5}$ K$^{-1}$ from 0.3 to 1000 K[2–5]. The magnitude, sign, and isotropy of its $\alpha$ makes $ZrW_2O_8$ especially attractive for applications.

While it is accepted that unusual low energy dynamics must be critical to the occurrence of NTE in $ZrW_2O_8$, the mechanism is not well understood in the sense that different proposed scenarios emphasize fundamentally different types of motion in the low energy dynamics [1,4,6,7]. In part, this may be reflective of the complexity of the underlying lattice [2, 8] which, though simple cubic, has a 44 atom basis and an unusual structure with bonds of abnormal length and large empty spaces [8]. Previous simulations [4] as well as XAFS experiments [6] have suggested that the low energy dynamics can be described in terms of the motion of nearly rigid $WO_4$ tetrahedra. Such a description is qualitatively consistent with the simple notion of $ZrW_2O_8$ dynamics driven by rigid unit modes [4]. The weakness of interactions between rigid units in $ZrW_2O_8$ can be represented in terms of a Maxwellian counting argument which finds that such motion is underconstrained compared to most minerals. Underconstraint of the $WO_4$ motion suggests that, despite the structural complexity, the low-energy dynamics can be reduced to understanding the relative motion of tetrahedra which exhibit libration and translation and are coupled by $ZrO_6$ octahedra.

In order to simulate the effects that an unusual degree of underconstraint might have on a structural system with relevance to $ZrW_2O_8$, we consider minimal models that have similar symmetry, but differ in their degree of constraint. These models evoke the essential motion of $WO_4$ rigid units and, for simplicity, are two-dimensional (2-D). Our work significantly extends earlier work that suggested the relevance of underconstrained 2-D models to NTE[9,10], and is complementary to earlier simulations which focused on the existence of zero-frequency rigid-unit modes (RUMs) occurring at special k values in the limit in which all interactions between polyhedra vanish[4]. Limitations of the RUM method have been discussed previously by Tao and Sleight[11].

Using calculations of the phonon spectra of our 2-D model systems, we show that there exist low energy phonon manifolds that soften with decreasing volume (thus generating NTE) *across the Brillioun zone* and which are a direct result of decreased Maxwellian constraint. Thus, this model reproduces the unique feature of $ZrW_2O_8$, namely the absence of a soft mode structural transition, despite the presence of extremely large anharmonic motion. The eigenvectors of this soft manifold exhibit a complex character in which translational and librational motion are inextricably mixed. We propose that the mixing observed in our 2-D models is a key signature of underconstraint, and directly related to the observation of unexpected dipole activity in far-infrared spectra [12], as well as recent neutron scattering results [7] of 3D $ZrW_2O_8$.

In Fig. 1 we show a structure consisting of a 2-D simple cubic Bravais lattice with a 3-atom basis. This is the reference constrained system to which we will contrast our results from underconstrained model systems. This structure is composed of simple, 2-D square elements, each with a central atom, four bonds of equal length and four internal angles of 90 degrees. These are the basic elements of our models and are analogues of the $WO_4$ tetrahedra in $ZrW_2O_8$. (Such a model might correspond to a material $XO_2$ with X a tetravalent ion) In our model all bonds are dynamic, however it is informative to consider the number of constraints that would be required to make each square rigid and to compare that to degrees of freedom. This is our Maxwellian constraint analysis. Fixing the nearest-neighbor bond lengths and the internal angles requires 7 constraints: 4 associated with bond length and 3 with bond angles. A 2×2 plaquet of this structure thus has 4×7 = 28 constraints compared to 32 degrees of freedom, D. An L×L plaquet has $6L^2 + 4L - 4$ constraints[13]. It has $D = 6L^2 + 4L$ arising from $3L^2 + 2L$ atoms. The value of D exceeds the number of constraints by exactly four, even as the number of atoms in the system becomes infinite. This means that there are only four eigenmodes which respect all bond length and angle constraints in this system. Three of these correspond to translation and rotation of the entire lattice. The fourth involves an alternate twisting of the squares[9, 10], and is regarded as relevant to NTE [6].

To introduce underconstraint, we insert linking atoms between the squares as shown in Fig. 2. (Such a model might correspond to a material $XY_2O_4$ where X and Y and tetravalent and divalent ions respectively.) For an L×L plaquet of this structure the number of constraints is $C = 13L^2 - 6L$ and $D = 14L^2 - 4L$. Thus, D exceeds the number of constraints by $L^2$. This reflects the existence of a *thermodynamic number of eigenmodes which preserve all bond angle and length constraints*. In this critical sense, this system is *macroscopically underconstrained*.

Another macroscopically underconstrained model system (not shown) can be obtained from the structure of figure 1 by removing the large atom from every 4th unit cell. This is the structure whose potential relevance to NTE has been suggested by Simon and Varma [9] and by Pantea at al. [10]. The constraint analysis of this system leads to $C = 21L^2 - 14L$ and $D = 22L^2 - 12L$. The difference residing in the $L^2$ term indicates a degree of underconstraint comparable to that of the system shown in figure 2. In addition, we have performed the mode analysis, as presented below, on this system and find similar results.

Analyzing the dynamics of these models allows us to make the connection between structural under-constraint, low-energy dynamics, and NTE. For this analysis, we used O and W masses for the small and large atoms respectively, an atom-atom separation of 1Å, and harmonic interatomic potentials defined by positional and angular force constants, k = 200 N/m and α = 20 N/m respectively. These energies are chosen to closely mimic the microscopics and low energy spectra found for 3D $ZrW_2O_8$. A weak angular force constant centered on the atoms at the corners of the squares is used to stabilize the structure or, alternatively, the system may be placed under tension, a circumstance that may be relevant to $ZrW_2O_8$ [14]. Phonon dispersion relations and eigenvectors are calculated from linearized equations of motion using Bloch eigenstates and diagonalizing q-dependent dynamic matrices of size 6x6, 14x14 and 22x22 respectively. Anharmonicity in these models comes via the angular forces and in some cases the inherent anharmonicity associated with the transverse motion of a system under tension[15]

In Fig. 1 calculated dispersion relations are shown for the *constrained* reference system. For this model system under tension there is a softening of the π-π transverse acoustic (TA) phonon with decreasing volume. The eigenvector for this mode corresponds to alternate twisting of squares. This is a "rigid unit mode" (RUM) [4] involving purely librational motion, and it is the mode mentioned earlier as the only bond-length and angle preserving eigenmode for this system. The softening of this mode can generate negative thermal expansion, however its influence is modest except very close to its soft mode transition.

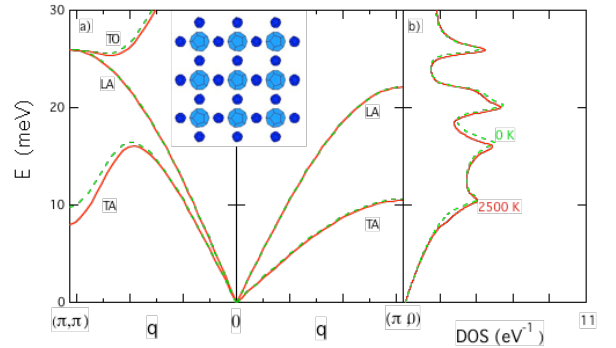

FIG. 1: (color online) a) Phonon dispersion relations for the lowest modes of the 2-dimensional structure shown are presented for the π-π and π-0 directions for lattice parameter values of 0.203 nm (dashed lines) and 0.202 nm (solid lines), respectively. Mode softening occurs only in the vicinity of π-π. b) Phonon density of states (DOS) shown for the same parameter values shows very little change. Temperatures refer to a free energy minimization calculation discussed in the text.

Underconstrained structures exhibit fundamentally different dispersion relations, which are shown in Fig. 2. Along the π-0 direction both the lowest transverse optic (TO) and the transverse acoustic (TA) phonon branches soften. Calculations of the full 2-D dispersion indicate that the softening of the TA phonons occurs only along the π-0 direction, however, the lowest TO branch softens with decreasing volume for all q. The concept of a special "NTE mode" at a particular q-value is thus not relevant here. Indeed, the logarithmic derivative, $\partial \ln \omega / \partial V$ (related to the Grüneisen parameter [1]), becomes essentially independent of q at low pressure indicating that the modes of this softening mode manifold each contribute equally to NTE.

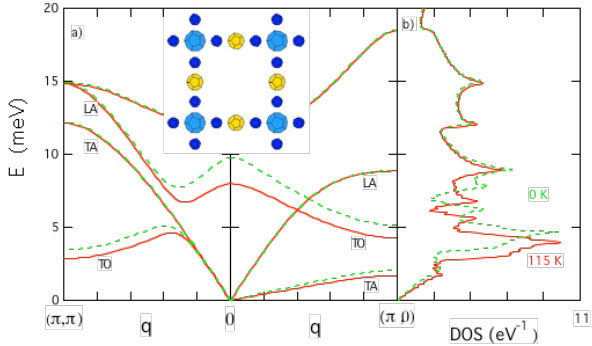

FIG. 2: (color online) a) Phonon dispersion relations for the lowest modes of an underconstrained structure are presented for the π-π and π-0 directions. Lattice parameter values are 0.406 nm (dashed lines) and 0.404 nm (solid lines), respectively. The lowest TO phonon branch softens with decreasing lattice size. b) Phonon density of states (DOS) for the same parameter values show a shift of spectral weight to lower energy with decreasing lattice size. Temperatures refer to a free energy minimization calculation at fixed pressure.

The existence of an extensive manifold of softening modes extending across the Brillioun zone generates a large volume-and-temperature dependent density of states (DOS) shown in Fig. 2b. The DOS peak at about 4.9 meV associated with this TO branch shifts downward significantly as the temperature increases from 0 to 115 K. This result is obtained by allowing the system to relax to the size that minimizes its Gibbs free energy, G = U - TS + pV, at constant pressure. In the example shown, a starting pressure of -0.89 N/m produces an NTE with a linear expansion coefficient of about α = -8.5 x 10$^{-5}$ K$^{-1}$. It is the influence of this volume dependent DOS, associated with a thermo-dynamically extensive number of softening phonon modes, that drives the thermal contraction with increasing temperature. Thus we see here an example of spectral weight downshift, a concept introduced for geometrically frustrated magnets[16,17] with suggested relevance for ZrW$_2$O$_8$ [18,19].

In Figure 3 we show the nature of selected modes of the softening mode manifold. At special points, e.g., 0-0 and π-π, relatively simple twisting motions are observed. Away from these special points, much more complex motions emerge in which twist and translation are inextricably mixed (Fig 3c). The infinite variety available via this complex mixing is critical to the existence of a thermodynamically infinite manifold of softening modes in underconstrained systems [20].

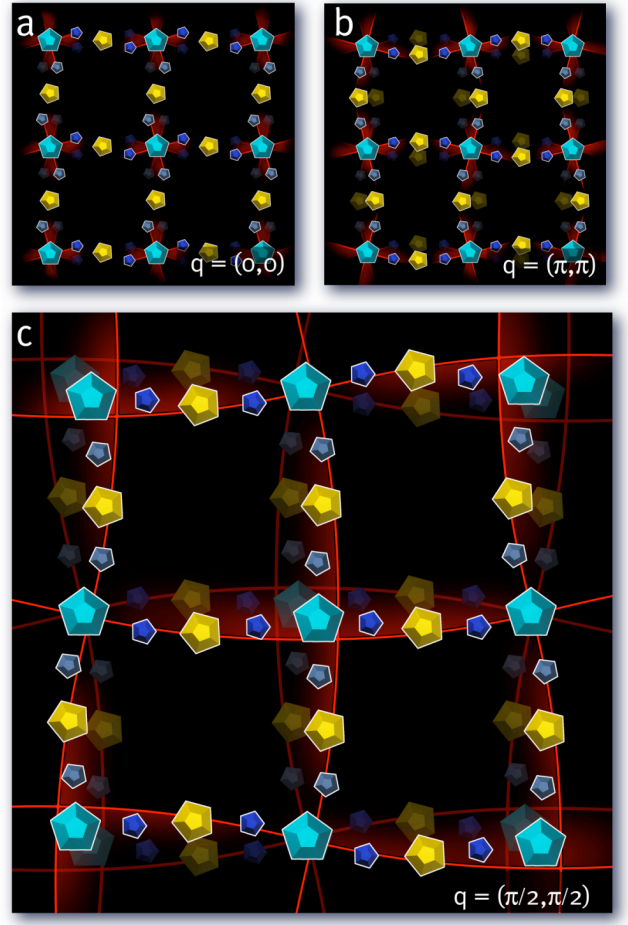

FIG. 3: Modes of the softening manifold are illustrated for a) q= (0,0), b) q= (π,π) and c) q= (π/2,π/2). The complex motion shown in c) is characteristic of ordinary points in the interior of the Brillioun zone and contrasts with the simple twisting motions at special points. Animations of these modes are available in the supplemental materials for this letter. (Illustration by Amadeo Bachar.)

For ZrW$_2$O$_8$, a sense of diminished constraint comes from the existence of terminal oxygen, commonly referred to as "unconstrained oxygen" [18]. Three of the oxygen atoms at the corners of each tetrahedron are in two-fold coordinations (providing linkages to Zr atoms). The 4th oxygen, however, is in a highly unusual one-fold (terminal) coordination and provides no linkage. This unusual structural aspect leads to substantially diminished constraint (relative to what would be expected in a more conventional or a hydrated structure). It is widely believed that the unconstrained terminal oxygen plays an important role in NTE – our modeling suggests a route to establishing this connection.

Mixing of twist and translational motion in the low energy dynamics of $ZrW_2O_8$ has been inferred from infrared phonon studies [12] and from recent neutron scattering measurements [7]. The dispersion calculations and eigenvector analysis presented here indicate that this is a fundamental signature of underconstraint.

One of our approaches to modeling uses a softening angle bond centered on the outer atoms of the squares; the other includes a negative pressure which stabilizes the structure. An inference of negative pressure in $ZrW_2O_8$ has emerged from a neutron scattering study of bond angles as a function of pressure by Jorgensen et al.[8] who noted "the surprising result that...volumes that correspond to negative pressure" arose naturally in their distance-least-squares-refinement analysis. It is interesting to note in this context that there are Zr-O bonds which are unexpectedly long in this structure; transverse motion of the oxygen atoms associated with these elongated bonds may play a critical role in the mechanism of NTE in $ZrW_2O_8$. These elongated bonds, which may be under tension due to steric constraints, are at the perimeter of an unusually large open space which lies opposite the underconstrained oxygen of the $W_1$ tetrahedra in the $ZrW_2O_8$ α structure (c.f., fig 2 of reference 8 and fig 1 of reference 7).

Unlike the simple model of Figure 2, real materials will tend to have longer range stabilizing forces which will remove the underconstraint formally. Our modeling indicates that as long as such forces are not overly strong, they do not effect the Grüneisen constants or NTE, but rather they establish a cut-off of the softening at a frequency scale below the physically relevant range. The dynamics and softening in the intermediate frequency range of particular relevance for NTE in real materials (3-10 meV) is essentially unaffected. Thus the aspect of underconstraint of importance for NTE, an infinite manifold of softening finite-energy modes, can survive the inclusion of stabilizing longer-range forces, which are surely present in real systems.

The modeling results presented here show how underconstraint can lead to a thermodynamic number of modes whose contribution to the entropy becomes the central driving force of a large NTE. Our modeling reveals both the relevance of a translational component in the low-energy dynamics, and, most importantly, the critical relationship between NTE and underconstraint. While lattice modeling provides valuable insights, it is also limited. Extensive mode softening leads to a region where the phonon entropy cannot be reasonably approximated by quasi-harmonic methods. This takes us outside the domain of ordinary Grüneisen parameter thermal expansion physics and into a region where it may be more appropriate to think in terms of Gibbs free-energy landscape flattening as the fundamental feature responsible for NTE. This view is suggested by infrared experiments which show extreme broadening of low energy phonon peaks and substantial spectral weight in an amorphous background at room temperature [12]. Approaches based on free-energy landscape flattening may provide a fruitful direction for further theoretical work which will allow a deeper understanding of NTE and its connection to frustration and underconstraint.

We acknowledge valuable conversations with Tobias Schultz, Dung-Hai Lee, Frank Bridges, Onuttom Narayan, B. Sriram Shastry, A. Peter Young and W.E. Pickett. We also acknowledge support from the NSF through grant DMR-0554796, and from NASA through UARC ARP award NAS2-03144.

[1] G. D. Barrera, J. A. O. Bruno, T. H. K. Barron, and N. L. Allan, J. Phys. Cond. Matt. 17, R217 (2005).
[2] T. A. Mary, J. S. O. Evans, T. Vogt, and A. W. Sleight, Science 272, 90 (1996).
[3] J. S. O. Evans, T. A. Mary, T. Vogt, M. Subramanian, and A. W. Sleight, Chem. Mater. 8, 2809 (1996).
[4] A. K. A. Pryde, K. D. Hammonds, M. T. Dove, V. Heine, J. D. Gale, and M. C. Warren, J. Phys. Cond. Matt. 8, 10973 (1996).
[5] A. W. Sleight, Current Opinion in Solid State and Materials Science 3, 128 (1998).
[6] D. Cao, F. Bridges, G. R. Kowach, and A. P. Ramirez, Phys. Rev. Lett. 89, 215902 (2002).
[7] M. G. Tucker, A. L. Goodwin, M. T. Dove, D. A. Keen, S. A. Wells, and J. S. O. Evans, Phys. Rev. Lett. 95, 255501 (2005).
[8] J. D. Jorgensen, Z. Hu, S. Teslic, D. N. Argyriou, S. Short, J. S. O. Evans, and A. W. Sleight, Phys. Rev. B 59, 215 (1999).
[9] M. E. Simon and C. M. Varma, Phys. Rev. Lett. 86, 1781 (2001).
[10] C. Pantea, A. Migliori, P. B. Littlewood, Y. Zhao, H. Ledbetter, J.C. Lashley, T. Kimura, J. Van Duijn, and G. R. Kowach, Phys. Rev. B 73, 214118 (2006)
[11] J. Z. Tao and A.W. Sleight, J.Solid State Chem. 173, 442 (2003).

[12] J. N. Hancock, C. M. Turpen, Z. Schlesinger, G. R. Kowach, and A. P. Ramirez, Phys. Rev. Lett. 93, 225501 (2004).
[13] For an internal square surrounded by 8 squares there are only 2 independent angular constraints, and hence a total of 6 constraints.
[14] Evidence for a sense of internal tension in $ZrW_2O_8$ comes in two forms: one is the experimental observation of an unusual bond-angle vs pressure relation that leads to the inference of negative pressure in fitting neutron date[8], the other is the unexpectedly long length of half of the Zr-O bonds.
[15] S.T. Thornton and J.B.Marion, Classical Dynamics of Particles and Systems (Brooks Cole, Pacific Grove, 2003)
[16] R. Moessner and J. T. Chalker, Phys. Rev. B 58, 12049 (1998).
[17] A. P. Ramirez and R. Moessner, Physics Today 59-2, 24 (2006).
[18] A. P. Ramirez and G. R. Kowach, Phys. Rev. Lett. 80, 4903 (1998).
[19] G. Ernst, C. Broholm, G. R. Kowach, and A. P. Ramirez, Nature 396, 147 (1998).
[20] Movies of these eigenmodes can be viewed in the auxilliary materials for this Letter.